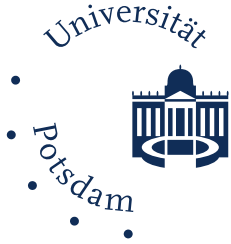
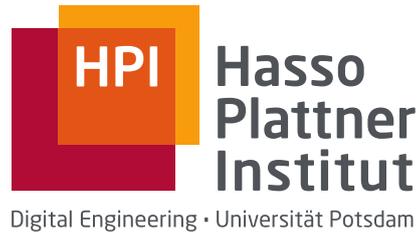

# Empirical Impact of Dimensionality on Random Geometric SAT

Empirischer Einfluss von Dimensionalität
auf das zufällige geometrische Erfüllbarkeitsproblem

## Flora Rädiker

Universitätsbachelorarbeit
zur Erlangung des akademischen Grades

**Bachelor of Science** *(B. Sc.)*

im Studiengang
IT-Systems Engineering

eingereicht am 2025-09-16 beim
Fachgebiet Algorithm Engineering
der Digital-Engineering-Fakultät
an der Universität Potsdam

| | |
|---|---|
| **Gutachter** | Prof. Dr. Tobias Friedrich |
| **Betreuer** | Dr. Andreas Göbel |
| | Nadym Mallek |

# Abstract

The Boolean Satisfiability Problem (SAT) is perhaps one of the most well-known problems in theoretical computer science. On the one hand, it is proven to be NP-complete, which means that it is generally considered hard to solve. On the other hand, the SAT problem has found many practical applications, which yield so-called industrial instances, and SAT solvers can often efficiently find solutions to such instances.

Closing this gap between theory and practice is a subject of current research. One approach is to identify properties of SAT instances that make them tractable. To aid in this, models for generating SAT instances have been proposed that mimic the properties of industrial instances. So far, attempts at creating such models were mostly unsuccessful, with instances being either too easy or too hard to solve, or missing important properties of industrial SAT instances.

In this work, we analyse a promising SAT model introduced by Giráldez-Cru and Levy [GL17] which is based on an underlying geometry. We empirically analyse the impact of this geometry's dimension on SAT instances with regard to three properties: the location of the satisfiability threshold, solver time, and size of proofs of unsatisfiability. Supplementing theoretical work by Bläsius *et al.* [Blä+23], we find that low-dimensional geometric instances are throughout very tractable. As dimension increases, instances from the geometric model seem to converge to hard uniform instances, which means that the geometric model is capable of representing the full range of easy to hard instances. We also observe that the satisfiability threshold in low-dimensional geometric instances occurs at lower densities. Additionally, low-dimensional instances behave very unlike uniform instances in that they have no hardness peaks in solver time at the satisfiability threshold. This coincides with proof size, which we show to not correlate to solver time at low dimensions.

# Zusammenfassung

Das Erfüllbarkeitsproblem der Aussagenlogik (SAT) ist vermutlich eines der bekanntesten Probleme in der theoretischen Informatik. Auf der einen Seite ist SAT bewiesenermaßen NP-vollständig, es wird also davon ausgegangen, dass es schwer zu lösen ist. Auf der anderen Seite hat SAT viele praktische Anwendungsfälle, aus denen sogenannte industrielle Instanzen hervorgehen, und SAT-Löser sind häufig in der Lage, Lösungen für solche Instanzen effizient zu finden.

Diese Lücke zwischen Theorie und Praxis zu schließen ist Gegenstand aktueller Forschung. Ein Ansatz besteht darin, Eigenschaften von SAT-Instanzen zu finden, die sie einfacher zu lösen machen. Hierfür werden Modelle zur Generierung von SAT-Instanzen entwickelt, die die Eigenschaften von industriellen Instanzen nachahmen. Bisher waren Versuche, solche Modelle zu entwickeln, größtenteils erfolglos, denn die generierten Instanzen waren entweder zu leicht oder zu schwer zu lösen oder ihnen fehlten wichtige Eigenschaften industrieller Instanzen.

In dieser Arbeit analysieren wir ein vielversprechendes SAT-Modell, das von Giráldez-Cru und Levy [GL17] vorgestellt wurde und eine zugrundeliegende Geometrie nutzt. Wir analysieren empirisch den Einfluss der Dimension dieser Geometrie auf SAT-Instanzen in Bezug auf drei Eigenschaften: den Ort der Erfüllbarkeitsschwelle, die Lösungsdauer und die Unerfüllbarkeitsbeweisgröße. Ergänzend zu theoretischer Arbeit durch Bläsius *et al.* [Blä+23] stellen wir fest, dass niedrigdimensionale geometrische Instanzen durchweg sehr leicht lösbar sind. Mit steigender Dimension scheinen Instanzen aus dem geometrischen Modell zu schwer lösbaren, gleichverteilten Instanzen zu konvergieren. Das bedeutet, dass das geometrische Modell in der Lage ist, die vollständige Bandbreite von leichten bis schweren Instanzen abzubilden. Wir beobachten auch, dass die Erfüllbarkeitsschwelle in niedrigdimensionalen Instanzen bei kleineren Dichten auftritt. Außerdem verhalten sich niedrigdimensionale Instanzen sehr anders als gleichverteilte Instanzen, denn sie zeigen bei der Erfüllbarkeitsschwelle keinen Schwierigkeitsgipfel in der Lösungsdauer. Passend dazu zeigen wir, dass die Unerfüllbarkeitsbeweisgröße im Falle niedriger Dimensionen nicht mit der Lösungsdauer korreliert.

# Contents

# 1 Introduction

Given a Boolean formula, the Boolean Satisfiability Problem (SAT) asks whether there exists an assignment of the formula's variables so that the formula evaluates to true. It is the first problem proven to be NP-complete [Coo71, Lev73] and it is frequently used in theoretical computer science to prove NP-hardness of other problems via reduction. Notably, one of the most famous open questions in computer science asks whether there exists a deterministic algorithm that solves any NP-complete problem, such as SAT, in polynomial time. Since no such algorithm is currently known (and is believed to not exist), the SAT problem is generally considered hard to solve. Despite this, SAT solving algorithms are often able to quickly find solutions to some instances. Consequently, the SAT problem has found practical applications in fields such as formal verification [Bie+03] and chip design [Heu+24, TN17], which yield instances known as *industrial* SAT instances.

To bridge the gap between easy instances in practice and theoretical hardness, one approach in current research is to identify properties inherent to certain SAT instances that make them tractable. To this end, models have been proposed that mimic the behaviour of industrial SAT instances. In the following, we first give an introduction to satisfiability, followed by a description of related research on random models. We then state our contributions.

## 1.1 Background on Satisfiability

To define what a SAT formula is, consider $n$ Boolean variables $x_1, ..., x_n$. Each of these variables may occur in its negated form, which is denoted as $\overline{x_i}$. We call a possibly negated variable a *literal*. For our purposes, multiple of these literals are disjunctively combined into a clause, for instance $x_3 \lor \overline{x_1} \lor \overline{x_4}$. A SAT formula is then made of a number of such clauses, which are conjunctively joined; for example:

$$(x_3 \lor x_1 \lor \overline{x_4}) \land (x_2 \lor x_3) \land (\overline{x_2} \lor x_3 \lor \overline{x_1} \lor x_4) \land (\overline{x_2} \lor \overline{x_4} \lor \overline{x_5})$$

To satisfy a SAT formula, it must hold that from each clause, at least one literal evaluates to true. The above formula, for instance, can be satisfied by setting $x_3 = \text{true}$ and $x_1 = x_2 = x_4 = \text{false}$.

# Chapter 1  Introduction

The *Handbook of Satisfiability* [Bie+09] gives an overview on the history of SAT, tracing its roots back to logic in ancient Greece. However, let us instead start in the 1970s, when Cook and Levin proved independently that SAT is NP-complete [Coo71, Lev73]. In this regard, a notable case is 2-SAT, where the formula's clauses always contain exactly two literals: It is proven to be deterministically solvable in polynomial time. Historically, research has been more and more focused on instances like these that consist of fixed-length clauses [FM09, p. 36]. Typically, the clause length is kept low, with 3-SAT especially gaining much attention, being the lowest clause length at which SAT is NP-complete.

### 1.1.1 Uniform SAT

As mentioned, various models for SAT have emerged. Given a set of parameters, these models generate random SAT instances, which can be used to assess the average behaviour of a parametrized SAT formula as well as measure SAT solvers' performance. The perhaps simplest of these models is *uniform SAT*. Given a number of variables, number of clauses, and clause length, the uniform SAT model draws an instance uniformly at random from all instances of the specified size. In spite of its simplicity, the uniform model can produce some of the hardest instances.

When clause length and number of variables are fixed, but the number of clauses is varied, one can observe changes in the behaviour of uniform SAT instances. Instances with few clauses compared to the number of variables tend to be satisfiable, while instances with many clauses tend to be unsatisfiable [HAB09]. The explanation for this is simple: Instances with few clauses are under-constrained, meaning that there are many satisfying variable assignments. On the other hand, increasing the number of clauses reduces the number of possible variable assignments, eventually leading to unsatisfiable instances.

The existence of these two extrema postulates that uniform SAT instances must undergo a transition from satisfiable to unsatisfiable. This *phase transition* of SAT has been a large topic of interest within the SAT community. There exists a so-called *satisfiability threshold*, at which SAT instances transition from satisfiable to unsatisfiable [Ach09]. At fixed clause length, it has been observed that the threshold's location is independent of instance size; it seems to be solely dependent on the ratio of the number of clauses to the number of variables, which is known as the *density* of a formula.

Some of the most interesting results in SAT are related to the satisfiability threshold. For one, hardness to solve seems to coincide with the threshold. The hardest uniform instances occur at the threshold, and the DPLL algorithm, which uses backtracking to solve SAT instances, provably requires an exponential number of steps to solve such instances [FP83]. In contrast, instances below or above the threshold are usually easier to solve. This effect was coined the "easy-hard-easy" pattern of SAT [HAB09]. Note that the term does not tell the full story: While instances below the threshold appear to be throughout benign and then abruptly increase in hardness at the threshold, the hardness seems to only gradually decrease above the threshold. [MSL92] offers an explanation for the "easy-hard-easy" pattern: The instances are hard at the threshold because they are neither under- nor over-constrained and thus only have few satisfying assignments or are unsatisfiable, which is hard to determine.

It should be noted that the above is largely based on empirical observations. Little is known for certain about the threshold; for instance its exact location. For 2-SAT, it is known to be at exactly density 1 [CR92], while there only exist empirical estimates at higher clause lengths [MMZ06].

It is also only conjectured that the satisfiability threshold is *sharp* [Ach09]. We call the threshold sharp if every instance below the threshold is a.a.s.[1] satisfiable, while every instance above it is a.a.s. unsatisfiable. Friedgut [Fri99] proved the existence of a sharp threshold which might depend on the number of variables. Recently, Ding *et al.* [DSS22] proved that the satisfiability threshold exists at fixed density and is sharp for large clause lengths.

## 1.2  Random Models for SAT

Unfortunately, the uniform SAT model does not adequately reproduce the properties of real-world industrial instances. For example, its instances are too hard to solve and, in real-world instances, the number of clauses in which a variable occurs is usually not evenly distributed. Thus, different models have been proposed that better mimic the properties found in industrial instances. These models generate random, parametrized SAT instances, which allows analysing the performance of SAT solvers based on these fine-grained parameters. This serves the purpose of gaining a deeper understanding of SAT as well as improving solver performance. Compared to industrial instances, parametrized models have the benefit of not relying on pre-existing instances with fixed properties [ABL09a]. Randomly generated instances are often used as benchmarks alongside industrial and handcrafted instances at the SAT Competition, an annual event held since 2002 in which SAT solvers compete against each other [SAT25].

---

[1] *asymptotically almost surely* (*a.a.s.*) refers to a probability tending to 1 as the number of variables goes to infinity

Ansótegui *et al.* [ABL09b] have observed that in many real-world SAT instances, variable occurrence follows a power-law distribution, meaning that few variables appear in many clauses while many variables appear in only few clauses. This property is known as *heterogeneity*. It has been incorporated in the *scale-free SAT* model [ABL09a] which features a parameter to control heterogeneity. However, empirical analysis has shown that SAT solvers still perform poorer on scale-free instances than on some large industrial instances [ABL09a, BFS19]. Moreover, Bläsius *et al.* [Blä+23] have shown that an unsatisfiable instance drawn from a power-law model likely requires superpolynomial time to solve. Thus, the scale-free model does not seem to sufficiently capture the properties of real-world SAT.

It has also been observed that real-world instances have *locality*, which means that variables which occur together in a clause are likely to share other clauses. This cannot be modelled by a power-law distribution, since the impact of a power-law is limited to the frequency of variables and does not influence their relationship. Instead, a natural choice is to model locality by an underlying *geometry*, which was first used by Bradonjić and Perkins [BP14]. This has been adopted by Giráldez-Cru and Levy [GL17] in a promising SAT model that incorporates a notion of geometry by placing clauses and variables of the instance on a geometric ground space and associating them based on their distance. In this way, variables which are closer together on the ground space will more likely share the same clauses. Additionally, variables can have weights assigned that follow a power-law distribution. As a result, this geometric SAT model is capable of modelling both a heterogeneous power-law distribution and locality. It is believed that this hybrid model can adequately model the spectrum of easy and hard SAT instances. In fact, Bläsius *et al.* [Blä+23] used a generalized version of the geometric model introduced in [GL17] to show that geometric instances are a.a.s. easy to solve. However, they assume constant dimensionality in the geometric ground space. We believe that the impact of dimensionality is of further interest on its own, since dimension likely influences properties such as satisfiability and hardness. We discuss this in the following section.

## 1.3  Quantifying Geometry in Geometric Graphs

SAT models are not the only area of research in which an underlying geometry and its dimension are of major interest. Similar observations with regard to the impact of geometry have been made in random graphs, which motivates our work on geometric SAT.

A classical random graph model is the Erdős–Rényi graph [ER59], in which each edge between two vertices exists with some fixed probability independently of other edges. It has been shown that various properties of Erdős–Rényi graphs also undergo a phase transition, quite similar to the phase transition found in uniform SAT. For example, depending on the number of vertices, there exists an edge probability above which the graph almost surely forms a giant component, while this is almost surely not the case below the threshold [ER60].

As with SAT, models have emerged that produce random graphs with various properties. Similar to real-world SAT instances, the degrees of vertices in real-world graph networks are often distributed according to a power-law. This can be represented by *Inhomogeneous Random Graphs* (IRGs) [BJR07], which are a generalization of Chung-Lu graphs [ACL01]. Equivalently to scale-free SAT, IRGs parametrize the vertex degrees by a power-law distribution.

It has also been observed that real-world graph networks have a notion of locality. Imagine, for instance, a friend graph: Two of one's friends are likely to be friends as well. Formally, this is known as the *clustering coefficient*, which measures the probability that two neighbours of a vertex are themselves adjacent. This property is taken into account by *Random Geometric Graphs* (RGGs) [DD23, Pen03], which—equivalently to the geometric SAT model—place vertices on a geometric ground space and draw edges based on some metric. Thus, vertices which are closer together on the ground space are more likely connected.

Finally, *Geometric Inhomogeneous Random Graphs* (GIRGs), introduced by Bringmann *et al.* [BKL19], encapsulate both IRGs and RGGs. They are based on a geometric ground space where vertices can additionally be assigned weights that follow a power-law distribution, just as weights can be assigned to variables in the geometric SAT model.

Properties of these graph models are currently much better understood than those of SAT, especially the impact of dimensionality in the two geometric models RGGs and GIRGs. For one, it has been proven that under certain circumstances, geometric graphs lose their geometry as dimension goes to infinity and they converge to a non-geometric model. Specifically, RGGs that use a spherical space provably converge to Erdős–Rényi graphs as dimension goes to infinity, given a large enough number of vertices [DD23, Theorem 1.(i)]. Also, GIRGs lose their geometric properties as dimension increases, and thus, if a torus is used as ground space, they provably converge to IRGs [Fri+24, Theorem 1.1]. Friedrich *et al.* [Fri+24] have also shown that for random geometric graphs to converge to their non-geometric counterpart, they require a homogeneous ground space, which both the sphere and torus fulfil. The hypercube, for instance, is not homogeneous. Because of its edges, distances between vertices on the hypercube are not independent. Thus, random geometric graphs on a hypercube do converge to some non-geometric model, but their edges do not become independent.

As can be seen, dimensionality has a large impact on models for geometric graphs. Even though geometric graphs share many properties with geometric SAT, no comparable research currently exists for geometric SAT models and the impact of dimensionality on its behaviour is largely unknown.

## 1.4  Our Contributions

We provide an empirical analysis of the impact of dimensionality on geometric SAT. We sample random SAT instances from the geometric SAT model used by Bläsius *et al.* [Blä+23], which is a generalization of the model introduced by Giráldez-Cru and Levy [GL17]. To analyse the impact of dimensionality at both low and high dimensions, we sample instances at up to 1 000 dimensions. We also sample instances from the uniform SAT model for comparison.

To better understand how dimensionality in the underlying ground space influences SAT instances and especially SAT solver's behaviour, we run different SAT solvers on our generated instances and record metrics on each instance (described in Chapter 3). Based on these metrics, we look at the following properties:

**Satisfiability**  We analyse the distribution of satisfiable instances and the location of the satisfiability threshold as well as its sharpness.

**Solver Time**  For each solver, we analyse the time it took to solve instances. We examine to what extent geometric instances exhibit the "easy-hard-easy" pattern known from uniform instances.

**Proof Size**  Some SAT solvers are able to output proofs of unsatisfiability for unsatisfiable instances; for a detailed description see Section 3.2.1. We analyse the size of those proofs, under the very general rule of thumb that a larger proof size indicates a harder-to-solve instance.

Recall that random geometric graphs converge to a uniform non-geometric graph as dimension goes to infinity. Due to the similarities of the RGG model to geometric SAT, we conjecture that the same is true for geometric SAT, as follows.

▶ **Conjecture 1.1**: **Convergence of Random Geometric SAT.** As dimension goes to infinity, the geometric SAT model converges to the uniform SAT model, with instances being indistinguishable at high dimensions.  ◂

Indeed, we empirically find that in all of the above properties, instances of geometric SAT gradually converge to the uniform case as dimension increases.

Additionally, based on the theoretical result that geometric instances are a.a.s. easy to solve [Blä+23] (see Section 4.1 for our reasoning), we conjecture that geometric SAT instances are much more tractable at lower dimensions and are more likely to be unsatisfiable, as follows.

▶ **Conjecture 1.2**: **Low-Dimensional Random Geometric SAT.** At low dimensions, instances from the geometric SAT model are much more tractable, they require smaller proofs, and they exhibit a satisfiability threshold at lower densities compared to uniform SAT.  ◂

Again, our findings support this conjecture. We observe that at up to 6 dimensions, SAT solvers do not exhibit the "easy-hard-easy" pattern on geometric instances. Instead, they are capable of solving all instances in a fraction of the time they require for uniform instances of the same size. At higher dimensions, instances do show a less pronounced hardness peak at the satisfiability threshold, slowly converging to that of uniform SAT.

We find that the satisfiability threshold occurs at much lower densities for low-dimensional geometric instances, but it reaches densities comparable (but not quite equal) to that of uniform instances at around 10 dimensions. We believe this to be due to the fact that the locality in geometric SAT instances leads to contradictions between variables which are closer together on the ground space (again, see Section 4.1 for our reasoning). Additionally, we find that the satisfiability threshold is much coarser at lower dimensions.

The shift in the satisfiability threshold and its coarseness have also been observed in the scale-free SAT model [BFS19], but they are overall less distinct than in our geometric model. This supports the current notion that geometric SAT is capable of representing a wider range of industrial SAT instances.

Comparing the impact of dimensionality on hardness and satisfiability, we find that the convergence of the threshold is faster than that of solver time. While the threshold reaches uniform-like densities quickly, there is still a noticeable difference in solver time between geometric instances at 250 dimensions and uniform instances.

Note that the theoretical result on hardness of geometric SAT by [Blä+23] is based on the proof size of an instance. Thus, we also analysed proof size. We find that low-dimensional geometric instances indeed require much smaller proofs. One would expect solver time to correlate at least somewhat to proof size, which is indeed the case for uniform instances. For geometric instances at up to 6 dimensions, however, we find that proof size does not correlate to solver time. In general, we identified the change from 6 to 7 dimensions to be the point at which geometric instances slowly start exhibiting properties known from uniform instances.

In the following, we introduce key concepts in Chapter 2, followed by a description of our experimental setup in Chapter 3. We state our results in Chapter 4.

# 2 Preliminaries

In this chapter, we formally define key concepts. We start by defining the SAT problem, which is at the heart of our work. There exist several variants of the SAT problem; our work uses $k$-SAT, in which (a) the clause length is always $k$ and (b) a formula is presented in Conjunctive Normal Form (CNF), i.e. it is made of conjunctively joined clauses.

▶ **Definition 2.1**: $k$-**SAT.** Let $x_1, ..., x_n$ be $n$ Boolean variables that can be either true or false. A $k$-SAT formula $\Phi$ is a conjunction composed of $m$ clauses $c_1 \wedge \cdots \wedge c_m$. Each clause is a disjunction of $k$ literals $l_1 \vee \cdots \vee l_k$. A literal represents one of the variables that may be negated, which is denoted as $\overline{x_i}$.

We call $\Phi$ *satisfiable* if and only if there exists an assignment of $x_1, ..., x_n$ such that $\Phi$ evaluates to true and we say it is *unsatisfiable* otherwise. ◀

We call $m/n$, which is the ratio of clauses to variables, the **density** of a SAT formula.

## 2.1 Random SAT Models

We introduce the two models that we use to generate random SAT instances. Our generation always produces an instance $\Phi$ with fixed $k$, $m$, and $n$.

Firstly, we define the uniform model, in which a SAT instance is drawn uniformly at random from all possible formulas.

▶ **Definition 2.2**: **Random Uniform** $k$-**SAT.** Let $\mathcal{U}_k(n, m)$ be a random uniform $k$-SAT formula with $n$ variables and $m$ clauses. We draw the formula's $m$ clauses independent of each other. The $k$ variables contained in each clause are drawn uniformly at random from all possible $\binom{n}{k}$ variables without repetition. Each literal's variable is then negated independently with probability $1/2$. ◀

Secondly, we generate SAT instances with an underlying geometry. As such, the model takes an additional, fixed parameter $d \in \mathbb{N}^+$ that specifies the dimension. The model is taken from [Blä+23].

Chapter 2    Preliminaries

Before we define the model itself, we first define the geometric ground space that provides the underlying geometry. Recall that Friedrich *et al.* [Fri+24] have shown that GIRGs require a homogeneous ground space to converge to a uniform graph model in which edges are independent. Thus, to achieve equivalent behaviour in our geometric SAT model, we choose a torus as ground space, which is homogeneous.

▶ **Definition 2.3**: **Geometric Ground Space.** Let $\mathbb{T}^d$ be the $d$-dimensional torus. Such a torus is a $d$-dimensional hypercube $[0, 1]^d$ in which opposite borders are identified, meaning that a coordinate of 0 is identical to a coordinate of 1. That is, we can impose coordinates $[0, 1)^d$.

To define Euclidean distance on the torus, let $p, q \in [0, 1)^d$ be two points on the torus. We define the circular difference between the $i$-th coordinates of $p$ and $q$ as

$$|p_i - q_i|_\circ = \min(|p_i - q_i|, 1 - |p_i - q_i|).$$

Then, the Euclidean distance between $p$ and $q$ is

$$|p - q| = \sqrt{\sum_{i \in [d]} |p_i - q_i|_\circ^2}. \qquad \blacktriangleleft$$

We use this definition of the $d$-dimensional torus to generate a geometric SAT instance, as follows.

▶ **Definition 2.4**: **Random Geometric $k$-SAT.** We say that a point $p \in [0, 1)^d$ on a torus $\mathbb{T}^d$ is random if each of its coordinates was drawn uniformly at random from $[0, 1)$.

To generate a geometric SAT instance, we sample one random point for each of the $m$ clauses and one for each of the $n$ variables. Then, we assign each clause the $k$ closest variables by Euclidean distance. Each literal's variable is negated independently at random with probability $1/2$. $\blacktriangleleft$

Figure 2.1 illustrates the generation of such an instance.

Note that we are only interested in analysing the properties of the geometric SAT model without a power-law distribution. As such, we leave out the variables' weights and temperature that are used in [Blä+23]. Additionally, we do not actively avoid repeating a clause when choosing negated literals (see Section 3.1.1).

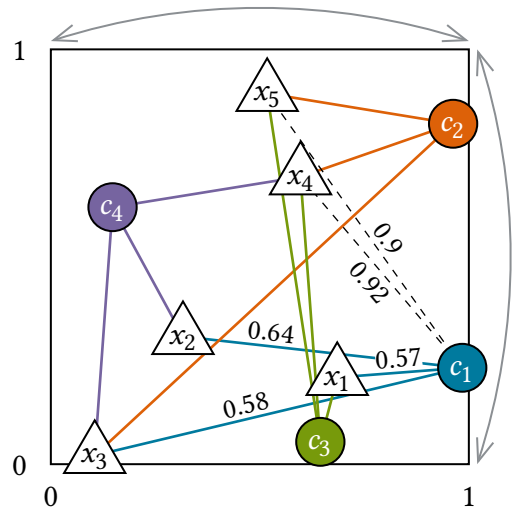

**Figure 2.1**: An example of the underlying geometry of a two-dimensional geometric instance. Note that the torus is shown flattened as a two-dimensional square; the arrows indicate identified borders. This instance consists of $n = 5$ variables $x_1, ..., x_5$ and $m = 4$ clauses $c_1, ..., c_4$. The three lines connected to each clause indicate its $k = 3$ nearest variables. Additionally, distances from clause $c_1$ to all five variables are shown.

## 2.2 Satisfiability Threshold

We say that an event occurs **asymptotically almost surely** (*a.a.s.*) if it occurs with a probability tending to 1 at $n \to \infty$.

As stated above, it has been observed that uniform instances undergo a sharp phase transition in which they change from being satisfiable a.a.s. to being unsatisfiable a.a.s. at increasing density. This is formalized in the following conjecture based on [Ach09].

▶ **Conjecture 2.5**: **Satisfiability Threshold Conjecture.** Let $\mathcal{U}_k(n, m)$ be a random uniform SAT instance with $k \geq 3$. Then, there exists a constant density $r_k > 0$ such that, for every $r > 0$ with $r \neq r_k$,

$$\lim_{n \to \infty} \Pr[\mathcal{U}_k(n, r \cdot n) \text{ is satisfiable}] = \begin{cases} 1, & \text{if } r < r_k \\ 0, & \text{if } r > r_k. \end{cases} \qquad \blacktriangleleft$$

$r_k$ is the location of the threshold, which we also call the **critical density**.

# 3 Experimental Setup

In this chapter, we describe the setup that we employed to achieve our results. To analyse random SAT instances, we conducted experiments in two steps:

1. Generate random SAT instances.

2. Solve those instances using a SAT solver and record various metrics for each instance.

For the first step, we developed a custom generation tool that we describe in Section 3.1. Note that we can perform the second step for different SAT solvers, incidentally yielding results on solver performance and a comparison of solver types. We describe our setup for this in Section 3.2.

We chose the Rust programming language [Rus] for the implementation of both generation and benchmarking. It allows us to write high-level code that is performant, memory-safe, and easy to read, ensuring that the correctness—especially of the generation code—is simple to verify.

## 3.1 Generation of Random SAT Instances

We developed a command-line interface that allows easy generation of uniform and geometric SAT instances. As input, it takes the type of instances to generate, a set of parameter combinations, as well as how many instances to generate per combination. As described in Definition 2.2 (Random Uniform $k$-SAT) and Definition 2.4 (Random Geometric $k$-SAT), both instance types take the clause length $k$, number of variables $n$, and number of clauses $m$ as parameters. Additionally, geometric generation takes a dimension $d$. The tool outputs one file per instance in the well-known DIMACS format, which simply stores the instance in CNF as a text file and is notably used by the SAT Competition [SAT11].

To accelerate generation, we employ several optimizations. Note that for uniform instances, clauses are generated independently of each other. Thus, we can efficiently write the output clause-wise without having to keep state between clauses. This is unfortunately not as easy for geometric instances, which is discussed in the following.

### 3.1.1 Efficient Generation of Geometric Instances

For geometric instances, as mentioned in Section 2.1, we do not avoid generating the same clause multiple times, which would require storing all previously generated clauses and thus immensely increase computation time and space. Avoiding clause repetition simplifies theoretical analysis [Blä+23], but its effect on empirical analysis is negligible since the number of distinct clauses increases with $n$.

Hence, as with uniform instances, geometric instance generation does not require keeping state between clause generation. However, generation of geometric instances does need the following initial setup step.

Recall that by Definition 2.4 (Random Geometric $k$-SAT), geometric instances are generated by placing points on a $d$-dimensional torus. Let points $v_1, ..., v_n \in [0, 1)^d$ represent each of the $n$ variables, and $c_1, ..., c_m \in [0, 1)^d$ represent the $m$ clauses. For each $c_i$, we now need to find the $k$ closest variable-points.

In a naïve approach, we would calculate the circular difference between each pair $c_i, v_j$ separately. Assuming that calculating the $d$-dimensional Euclidean distance takes time linear[2] in $d$, this calculation requires a run time of $\mathcal{O}(mnd)$. Then, for each clause $c_i$, we sort all variable-points by distance to $c_i$ and pick the $k \leq n$ closest, leading to a total run time of $\mathcal{O}(mnd + mn \log n)$. We could improve the $k$-closest selection by utilizing Introselect [Mus97]. It allows selecting the $k$-th smallest element of an array in linear time, thus providing a way to also find the $k$ smallest elements in linear time by comparison. This would reduce the total run time to $\mathcal{O}(mnd)$.

However, we can instead drastically improve performance by using a *$k$-dimensional tree*, as introduced in [Ben75]. It is well-suited for solving the all-nearest-neighbours problem, providing an average run time of $\mathcal{O}(\log n)$ for insertion as well as search.[3] With an initial setup step to insert the $n$ variable-points into the tree, which takes time in $\mathcal{O}(n \log n)$, the average run time to determine the $k$ closest variables for each clause thus becomes $\mathcal{O}(n \log n + mk \log n)$.

In practice, this means generation of large geometric instances scales much better. For example, we measured a 200-fold speedup for the generation of a $k = 3, n = m = 10^5, d = 3$ instance, taking around 350 ms on our hardware.

---

[2] We pragmatically expect that squaring floating point numbers takes constant time on real hardware.

[3] Note that [Ben75] assume constant dimension and their average search run time is empirical.

## 3.2 Benchmarking SAT Instances

In this second step, we let SAT solvers solve the instances that we generated previously and record the following metrics on each instance:

- whether the instance was found satisfiable or unsatisfiable,
- the time it took to solve the instance,
- proof size (see Section 3.2.1).

We chose instances small enough to be solvable within a timeout of 15 minutes. All solvers were run on an AMD EPYC 7742 processor at 2.25 GHz.

We chose three SAT solvers for our experiments (see Table 3.1). `minisat`, being a simple and very popular SAT solver, was chosen for its historical significance, winning the SAT Competition 2005. Second, we chose `glucose`, which is based on `minisat`, but adds its own features. Most notably for our work, it supports outputting proofs of unsatisfiability. Finally, we chose `kissat` as a current state-of-the-art solver. `kissat` and a related solver, `CaDiCaL`, were first in the Main Track of all SAT Competitions since 2020 [SAT25].

All three solvers are complete *conflict-driven clause learning* (CDCL) solvers, meaning that they solve both unsatisfiable and satisfiable instances. We opted against using incomplete *stochastic local search* (SLS) solvers, which are more efficient at the satisfiability threshold but only support satisfiable instances, since our main focus is on geometric instances regardless of satisfiability.

### 3.2.1 Proof Size

While a satisfying assignment found by a SAT solver can be trivially checked, unsatisfiable instances require a more sophisticated method of verification.[4] Some

Table 3.1: The SAT solvers used in our experiments.

| Solver   | Year | Version | Reference | Proof support |
|----------|------|---------|-----------|---------------|
| `minisat` | 2010 | 2.2.0   | [SE05]    | no            |
| `glucose` | 2018 | 4.2.1   | [AS09]    | yes           |
| `kissat`  | 2025 | 4.0.2   | [Bie+24]  | yes           |

---

[4] Otherwise, unsatisfiable instances can't get no satisfaction [Rol65].

# Chapter 3 Experimental Setup

SAT solvers are able to emit proofs of unsatisfiability, typically in the DRAT format, which is required by the SAT Competition in its Main Track [Heu+24, SAT23].

A proof in DRAT format consists of a list of clauses. Each of those clauses is either added to or deleted from the original SAT instance. The idea is to modify the SAT instance in such a way that it is at least as satisfiable as the original instance (with regard to satisfying variable assignments) at every step [BCH21]. While deletion of a clause always fulfills this trivially, addition of a clause must not violate this property.

As they try to solve an instance, SAT solvers learn clauses that can be added to the original instance. In fact, especially early solvers would simply output their learned clauses as a proof [SAT23]. Modern SAT solvers also instruct to delete irrelevant clauses in order to speed up verification.

After applying all clause additions and deletions, a proof verifier can check whether the proof is valid using *unit propagation*, which deletes and simplifies clauses by utilising information gained from single-literal clauses. Assume the modified instance contains a clause consisting of only the literal $x$. Then, to satisfy the formula, $x = $ true must hold. Thus, any clause containing $x$ is satisfied and can be deleted (except for the single-literal clause itself), while any occurrences of $\bar{x}$ can be removed since these can no longer contribute to satisfiability. If removal of $\bar{x}$ leaves the clause empty, this clause cannot be fulfilled, so the instance is proven to be unsatisfiable. To initiate this process, a proof always contains a single-literal clause.

In our experiments, we recorded the following metrics on all proofs of unsatisfiability:

- the total number of proof clauses,
    ‣ how many of those were additions or deletions,
- the total number of literals across all clauses,
    ‣ how many of those occurred in added or deleted clauses, and
- the maximum length of any proof clause.

Our analysis showed that the numbers of additions and deletions are mostly proportional to the total number of clauses or literals. Similarly, the total number of literals scales with the total number of clauses. Therefore, our further analysis is done only on the total number of proof clauses and the maximum length of any clause.

# 4 Empirical Results

In this chapter, we present our results. These were obtained by generating random instances, solving them using three SAT solvers, and then visualizing various properties of the instances.

We generated 3-SAT instances with fixed $n = 300$ variables, which is about the maximum instance size where solving is still feasible even for the hardest uniform instances. We varied the number of clauses $m$ from 300 to 3 000 to achieve densities $m/n \in [1, 10]$, which covers the important densities for both uniform and geometric instances. Geometric instances were generated with dimensions $d \in [1, 50]$ and additionally with specific higher dimensions up to $d = 1\,000$. For each distinct parameter combination, we generated 100 instances in order to average recorded metrics.

We also examined whether our results hold for instances with a different clause length or number of variables by generating few instances of other sizes. We discuss this at the end in Section 4.5.

## 4.1 Impact of Dimensionality

We will first explain our conjectures, then present our results. Recall the following conjectures from the introduction.

▶ **Conjecture 1.1**: **Convergence of Random Geometric SAT.** As dimension goes to infinity, the geometric SAT model converges to the uniform SAT model, with instances being indistinguishable at high dimensions. ◀

▶ **Conjecture 1.2**: **Low-Dimensional Random Geometric SAT.** At low dimensions, instances from the geometric SAT model are much more tractable, they require smaller proofs, and they exhibit a satisfiability threshold at lower densities compared to uniform SAT. ◀

To understand why we expect geometric instances to be substantially more tractable, recall that Bläsius *et al.* [Blä+23] have shown that geometric instances are a.a.s. easy to solve. Specifically, they proved that as $n$ goes to infinity, a geometric instance contains a.a.s. an unsatisfiable subformula of constant size, which can be found in $\mathcal{O}(n \log n)$ time.

Also recall that uniform instances with low density are likely satisfiable, while instances with high density are likely unsatisfiable. Instances at the threshold in-between are hard because they are neither under- nor over-constrained [MSL92]. It is likely that the locality prevalent in geometric instances influences how quickly the number of clauses can make an instance over-constrained since clauses that are located closer together on the torus likely share some or all variables, creating unsatisfiable substructures. Accordingly, we expect the satisfiability threshold in geometric instances to occur at lower densities. Additionally, small substructures directly imply that geometric instances require a smaller proof size, as found by [Blä+23]. We also expect solvers to find such proofs more quickly, since they are able to find smaller substructures more efficiently.

Figure 4.2 shows an overview of our results: It depicts the change in satisfiability threshold and solver time by dimension. Evidently, the critical density as well as solver time increase with dimension. Also, there is no noticeable difference between geometric instances at $d = 1\,000$ and uniform instances, neither in satisfiability nor in solver time. Many of the following figures display the same data, but with a different focus and finer granularity.

In the next sections, we dive into each of the three properties that we analysed.

## 4.2 Satisfiability Threshold

For uniform SAT instances, empirical results by [MMZ06] suggest that the critical density is located at $r_3 \approx 4.267$. Looking at Figure 4.3, we find that the satisfiability threshold generally occurs at much lower densities for low-dimensional geometric instances, being below $m/n = 2$ at $d = 1$. The critical density increases with each dimension, reaching $r_3$ fairly quickly, with $m/n \approx 4.1$ at $d = 7$. This is in accordance with our conjectures: Low-dimensional geometric instances are more likely to be unsatisfiable due to small unsatisfiable substructures induced by locality.

Additionally, the threshold appears to be much coarser at low dimensions, which has also been observed in scale-free SAT [BFS19] (see Section 1.4). At $d = 1$, instances with density $m/n = 1.0$ are mostly satisfiable and take until $m/n = 3.0$ to be mostly unsatisfiable. In contrast, at $d = 7$, the ratio of satisfiable instances decreases much quicker within a difference in density of about 0.6.

Satisfiability Threshold  Section 4.2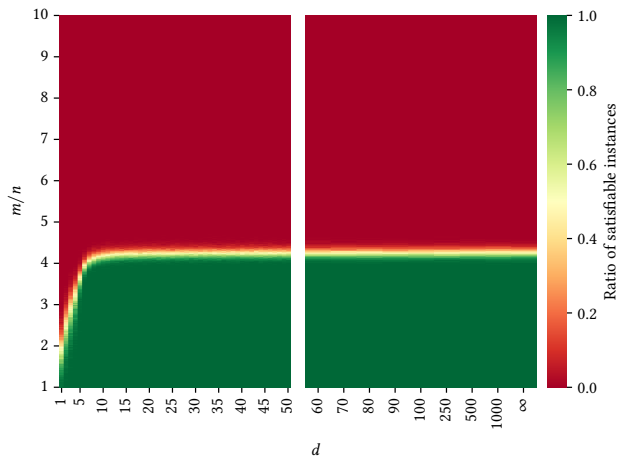

**(a)** Ratio of satisfiable instances.
Red areas contain only unsatisfiable instances,
while green areas contain only satisfiable ones.

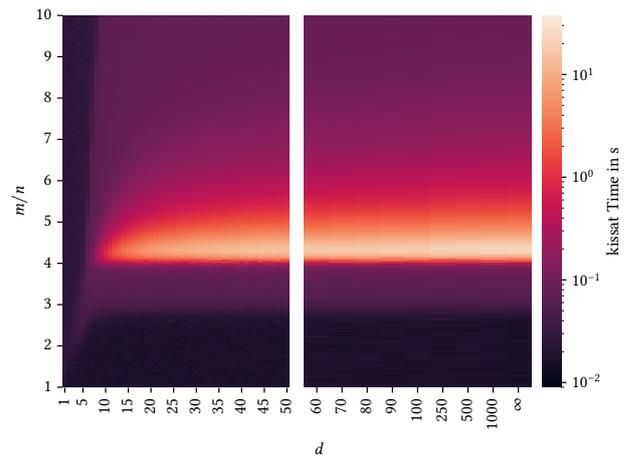

**(b)** Average time to solve by `kissat`.
Note the additional plateau
at around $10^{-1}$ s solver time.

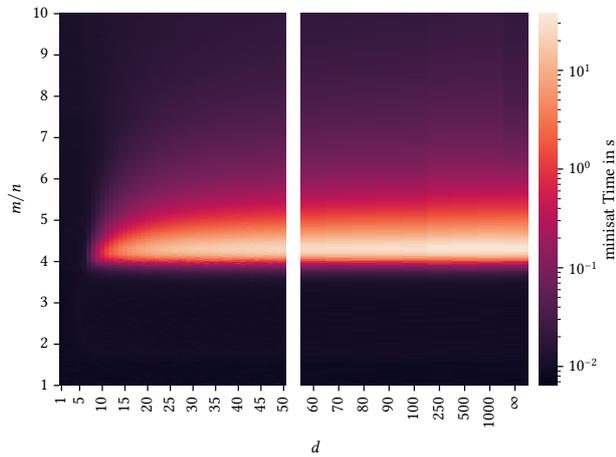

**(c)** Average time to solve by `minisat`.

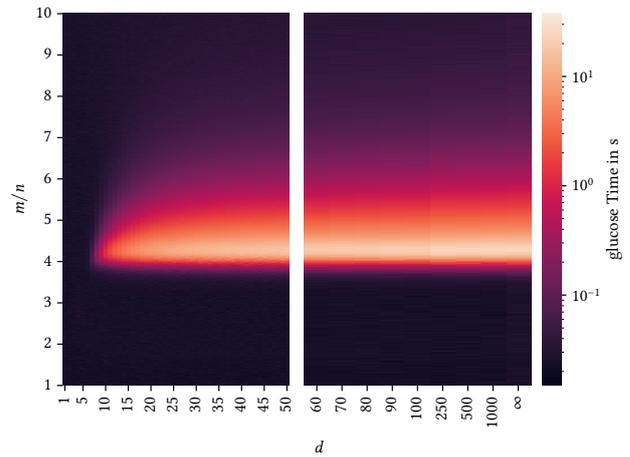

**(d)** Average time to solve by `glucose`.

**Figure 4.2**: Heatmaps showing the ratio of satisfiable instances (a) and the average solver time (b), (c), (d).

On the vertical axis, each plot shows the density $m/n$.

On the horizontal axis, each plot shows the dimension. The left area of each plot shows all dimensions $d \in [1, 50]$, while the right area shows non-equidistant dimensions $d \in \{60, 70, ..., 100, 250, 500, 1\,000\}$. Additionally, values for uniform instances are shown at the end for comparison ("$\infty$").

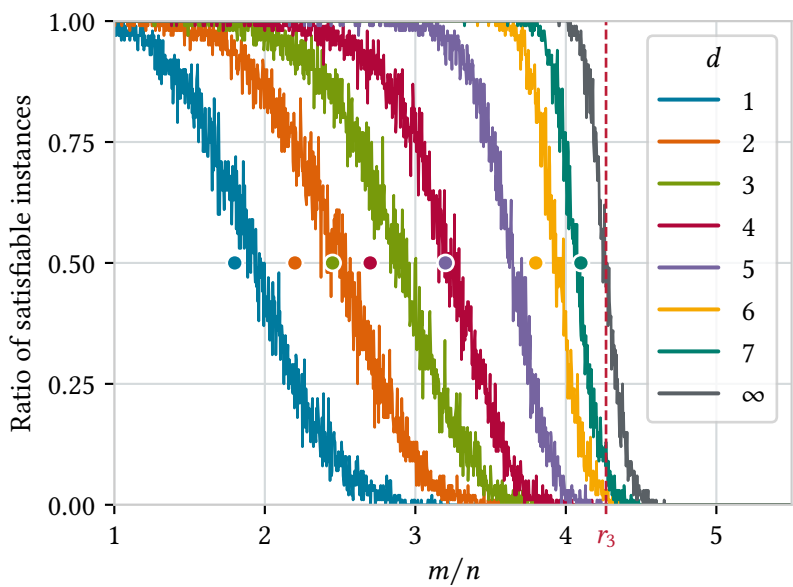

**Figure 4.3**: The ratio of satisfiable geometric instances by density at low dimensions $d \in [1, 7]$. Additionally, the ratio of satisfiable uniform instances is shown for comparison ("$\infty$"), which—as expected—reaches a ratio of $1/2$ at $r_3 \approx 4.267$ [MMZ06].

For each dimension, the circles show the density at which the plateau observed in `kissat`'s solving time reaches its maximum (see Figure 4.4). The circles are placed at ratio $1/2$ since, if the plateaus are related to the ratio of satisfiable instances, the maximum would be expected to occur at a ratio of $1/2$.

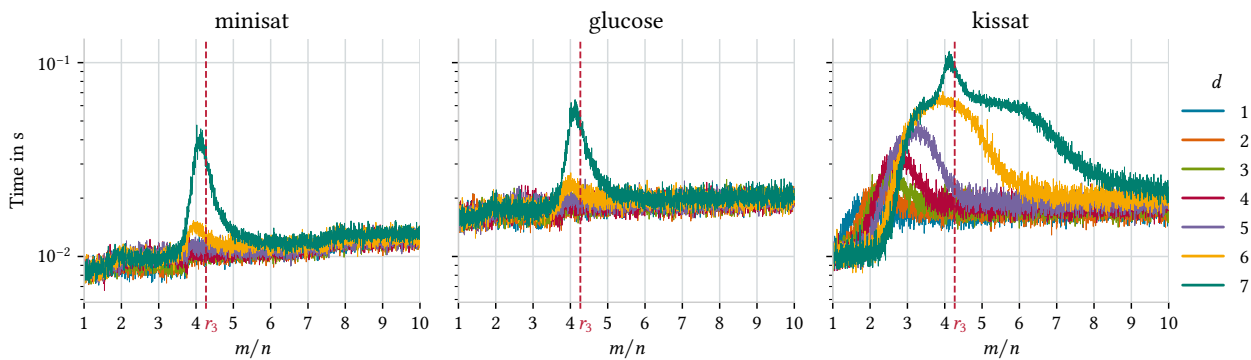

**Figure 4.4**: Average solver time at low dimensions up to $d = 7$ by density.

At $d = 7$, all solvers display a significant peak in solving time just below the uniform satisfiability threshold at $r_3 \approx 4.267$.

Unlike the other two solvers, `kissat`'s solving time exhibits a plateau that is present at every dimension.

## 4.3 Solver Time

Recall that uniform instances exhibit a so-called "easy-hard-easy" pattern: For fixed clause length *k* and *n* variables, they are easy to solve at low densities, then sharply become hard to solve at the satisfiability threshold before slowly becoming more tractable at higher densities. This is clearly evident in the—as we like to call them—"sunset plots" from Figure 4.2 showing solver time for three solvers. At higher dimensions, solver time sharply increases around $r_3 \approx 4.267$ and only slowly decreases again. Overall, all three solvers behave very similar.

To facilitate a more fine-grained analysis, Figure 4.4 displays solver time specifically at low dimensions. At these low dimensions, `minisat` is fastest, followed by `kissat` and `glucose`. In contrast, Figure 4.5 shows solver time for uniform instances: `minisat` still performs best below and above the uniform satisfiability threshold, while `glucose` and `kissat` perform better at the hardest instances around the threshold. `kissat` is not only fastest over all uniform instances, but in sum also takes the shortest time to solve all instances that we generated. This is in spite of kissat performing worse than glucose and minisat at higher densities, and worse than minisat at lower densities. kissat is especially fast at the satisfiability threshold of uniform or high-dimensional instances. A possible explanation could be that kissat, being the most recent solver, is highly optimized to solve hard instances around the critical density, at the cost of taking slightly longer for easier instances.

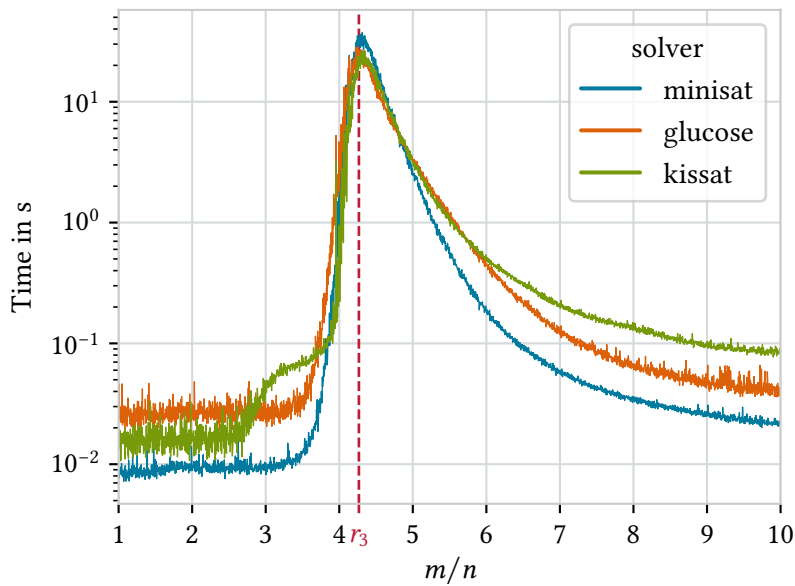

**Figure 4.5**: Average solver time of uniform instances by density. As expected, solvers perform worst near the critical density $r_3 \approx 4.267$. At the critical density, `kissat` marginally performs best, followed by `glucose` and lastly `minisat`. Note that because of this, `kissat` is the fastest solver in total over all uniform instances, even though the logarithmic time axis might suggest otherwise.

Interestingly, low-dimensional geometric instances do not seem to exhibit an "easy-hard-easy" pattern; all instances seem to be fairly benign. This can be observed in Figure 4.2 as well as, in greater detail, in Figure 4.4. It is evident that density does not influence glucose's solver time much up to $d = 6$. It looks similar for minisat, where solver time does increase slightly with density, but the first visible peak occurs only at $d = 6$. Both minisat and glucose show a large peak in solver time only from $d = 7$. The peaks at $d = 7$ occur just below the critical density of 3-SAT. As can be seen in Figure 4.2, the peak in solver time correlates to the shift in critical density, with the density at which the peak occurs increasing at higher dimensions.

Likewise, kissat displays the well-known "easy-hard-easy" peak at $d = 7$. However, contrary to glucose and minisat, kissat's solving time exhibits a large plateau that seems to persist independently of the satisfiability threshold. This plateau is present regardless of dimensionality as can be seen in Figure 4.2 (b). For high-dimensional and uniform instances, `kissat`'s solving time abruptly increases at $m/n \approx 3$. At low dimensions, this plateau occurs at lower densities, and solving time decreases again at higher densities. To examine a possible relation to the usually observed "easy-hard-easy" pattern, we added the densities at which `kissat`'s plateau reaches its maximum to Figure 4.3. If the plateau is related to satisfiability, we would expect its maxima to correlate to the satisfiability threshold. As can be seen, there does not seem to be an immediate connection to the threshold, suggesting that `kissat` exhibits another "easy-hard-easy" pattern unrelated to the satisfiability threshold. We do not currently have an explanation for this observation.

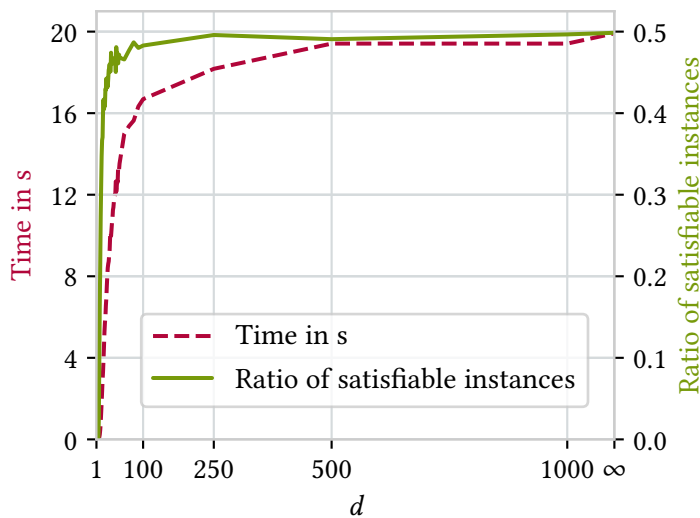

**Figure 4.6**: Comparison of solver time and satisfiability ratio by dimension at the uniform SAT threshold. Each value is averaged over all instances with density $m/n \in [4.20, 4.33]$.

In Figure 4.6, we visualize the shift in solver time and satisfiability ratio as dimension increases by looking at instances specifically around the critical 3-SAT density $r_3$. At $d = 100$, there is still a noticeable difference in solver time between geometric instances and uniform instances. This is in stark contrast to the satisfiability ratio, which reaches uniform-like values much more quickly. It is clearly evident that while the satisfiability ratio reaches values near the uniform ratio at around $d = 100$, solver time takes until around $d = 500$. This is particularly noteworthy since it suggests that there exist SAT instances where the satisfiability threshold behaves similarly to the uniform case, while still being easier to solve.

## 4.4 Proof Size

For each instance, we also recorded metrics on the proofs supplied by `glucose` and `kissat`, as described in Section 3.2. This gives valuable insight into real solver behaviour.

Note that the work done by [Blä+23] is based on an upper bound on the proof size of geometric SAT. They showed that, at large $n$, there exists asymptotically almost surely a subformula consisting of at least $2^k$ clauses which all contain the same variables. Since literals are negated at random, this likely causes the formula to be unsatisfiable. Due to the constant size of the subformula, proofs can be found quickly, making the instances tractable. Analysing real-world solver proofs can thus strengthen this result, perhaps providing evidence even for smaller $n$.

Also note that recording solver proof size measures the actual proof size required by the solver, not the theoretical upper bound. Solvers' optimizations usually focus on solving speed, not proof size beyond reasonable means (like the SAT Competition's timeout on proof checks). As such, proof size should not be used to compare solvers.

Our findings suggest that geometric instances indeed require smaller proofs, which is discussed in the next section. Apart from that, low-dimensional proofs behave very differently from uniform instances' proofs with regard to solver time, which is discussed in Section 4.4.2.

### 4.4.1 Impact on Proof Size

Figure 4.7 displays two metrics of proof size: the number of proof clauses and the maximum length of a proof clause. Note that our real-world number of proof clauses is greater than or equal to the well-known theoretical *resolution size*, which measures the minimum number of clauses any proof of unsatisfiability requires. In the same manner, the real-world maximum length of a proof clause is greater than or equal to the *resolution width*, which measures the smallest maximum length of a clause that any proof requires.

It immediately becomes clear that low-dimensional instances require comparatively smaller proofs. Up to $d = 6$, both `kissat` and `glucose`'s proofs have comparatively small maximum proof clause lengths, requiring less than 10 literals on average. Of the two, `kissat` outputs especially small clauses at low dimensions, either two or three literals at $d = 1$ (Figure 4.9). This is not surprising since most low-dimensional instances contain contradictions between only two or three variables, which are thus the only variables in the relevant part of a solver proof. Listing 4.8 shows an example of such a real-world solver proof. It is evident that it suffices to show a contradiction between two variables, although `kissat` outputs much more clauses containing irrelevant variables. For some other instances, `kissat` immediately finds a simple contradiction and outputs only a small proof of less than 10 clauses. We assume this is due to `kissat` not finding a refutation immediately, in which case it will additionally output any clauses it has learned so far. Unsurprisingly, 19 out of 21 clauses which contain the two refuting variables also contain the same third variable due to geometric locality leading to extremely similar clauses at $d = 1$. Since each clause's literals are negated uniformly at random, it is very likely that an instance contains clauses which consist of the same variables but are differently negated. At high enough densities, i.e. if there

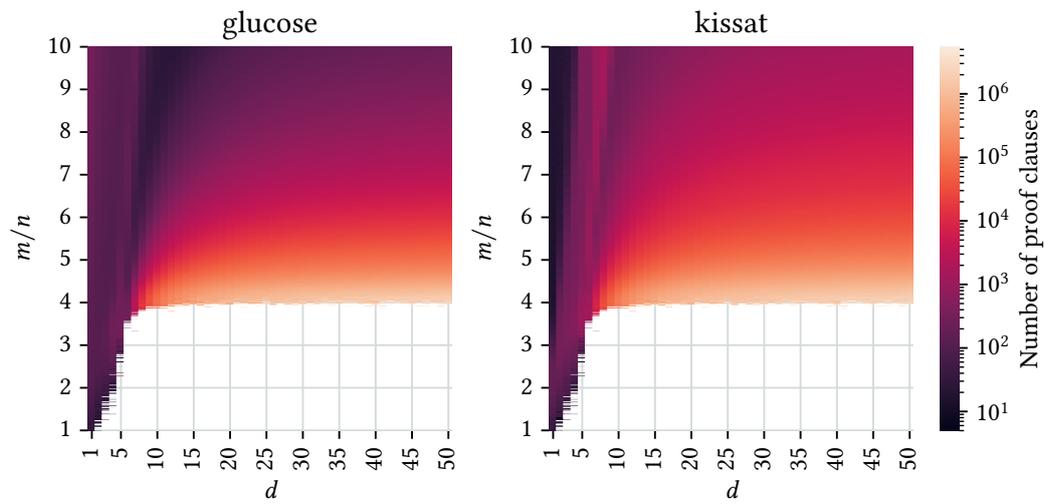

(a) Average number of added or deleted clauses that the proof is composed of.

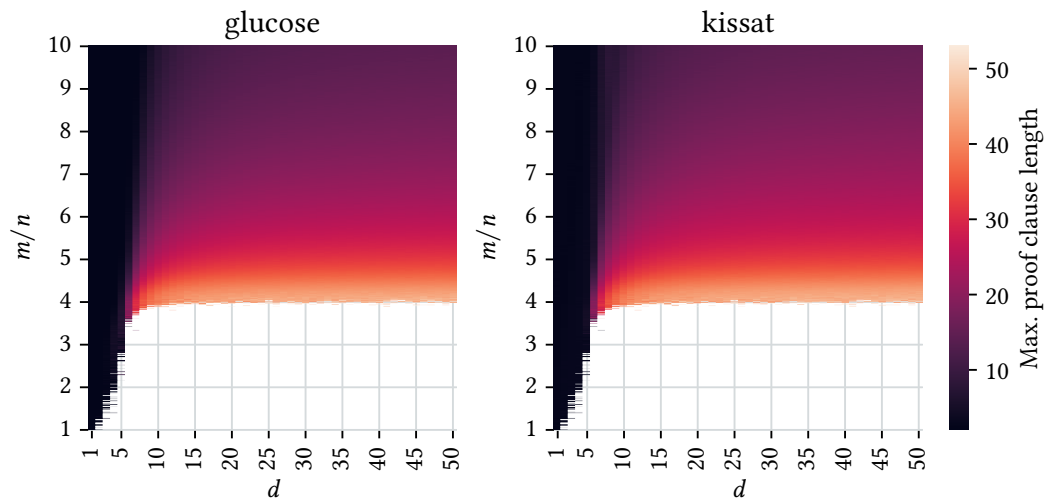

(b) Average maximum length of any clause in the proof.

**Figure 4.7**: Two metrics of proof size by solver.

Each plot shows the metric of proof size by dimensions up to $d = 50$ and density. In the empty regions at low densities, all instances are satisfiable and thus have no proofs.

Note that since the heatmaps show averages, information on few instances that require larger proofs is lost; refer to Figure 4.9 for details.

$x_{28} \lor x_{42} \lor x_{281}$

$\overline{x_{28}} \lor \overline{x_{42}} \lor x_{281}$

$x_{28} \lor x_{276} \lor x_{281}$ (×3)

$x_{28} \lor x_{276} \lor \overline{x_{281}}$

$x_{28} \lor \overline{x_{276}} \lor x_{281}$ (×3)

$x_{28} \lor \overline{x_{276}} \lor \overline{x_{281}}$ (×5)

$\overline{x_{28}} \lor x_{276} \lor x_{281}$

$\overline{x_{28}} \lor x_{276} \lor \overline{x_{281}}$

$\overline{x_{28}} \lor \overline{x_{276}} \lor x_{281}$

$\overline{x_{28}} \lor \overline{x_{276}} \lor \overline{x_{281}}$ (×4)

$x_{281} \lor x_{28}$

$x_{281} \lor \overline{x_{28}}$

$\overline{x_{281}} \lor x_{28}$

$\overline{x_{281}} \lor \overline{x_{28}}$

$x_{28}$

**(a)** Excerpt of some of the instance's clauses. All clauses that contain either of the refuting variables $x_{28}$ or $x_{281}$ are shown, with some occurring multiple times.

**(b)** The last five clauses (all are additions) of `kissat`'s proof. The proof consists of 53 clauses in total, but only these last five clauses reference the refuting variables $x_{28}$ and $x_{281}$.

**Listing 4.8**: Excerpt of an instance (a) and excerpt of the proof of unsatisfiability for this instance by `kissat` (b). The instance has parameters $k = 3$, $n = 300$, $m = 1\,702$, $d = 1$.

As an effect of geometric locality, there exists no clause containing only one of the refuting clauses $x_{28}$ and $x_{281}$; they always appear together. Also, the two refuting variables are always accompanied by one of two other variables: $x_{42}$ or $x_{276}$, with $x_{276}$ occurring in 19 out of 21 clauses.

Although `kissat`'s proof consists of 53 clauses in total (both additions and deletions), only the last lines shown here reference $x_{28}$ or $x_{281}$, suggesting that `kissat` outputs its learnings before finding a refutation.

exist enough clauses, this almost surely causes a conflict. This is in accordance with the theoretical findings by Bläsius *et al.* [Blä+23].

Proofs are especially large for hard instances at the uniform satisfiability threshold, but only beyond $d = 6$. This is very similar to the increase in solver time, and it seems to coincide with our observation that the first major increase in solver time occurs at $d = 7$. As with solver time, proof size displays a "hard-easy" pattern, where unsatisfiable instances, as soon as the satisfiability threshold is reached, require very large proofs, which only gradually decreases again at higher densities. One can even argue that, again, `kissat` displays a "second threshold": At dimensions up to $d = 6$, `kissat` requires more proof clauses at lower densities, and after a threshold that shifts up proportionally to dimension, the number of proof clauses is lower. In contrast, `glucose`'s average number of proof clauses is quite steady at low dimensions.

### 4.4.2 Proof Size Versus Solver Time

It would be reasonable to assume that a larger proof correlates to harder instances since finding a larger proof also requires more work by the solver. In fact, some solvers simply output their learned clauses as proof [SAT23], in which case it would be expected that proof size at least correlates to search tree depth. As a means to compare solver time to density, we plotted all instances of select dimensions as a scatter plot, with solver time on one axis and proof size on the other. The result can be seen in Figure 4.9. The plot gives various information on solver behaviour and geometric instances. We start by investigating solver behaviour, then discuss geometric instances.

Much like in Figure 4.7, `kissat` and `glucose` behave very similar with regard to maximum proof size, but show some differences at the number of proof clauses. At higher dimensions, both display a similarly strong correlation between solver time and proof size, as one would expect. At lower dimensions, `kissat` is again very good at producing small proofs for a majority of instances, while `glucose` requires overall longer proofs. For both solvers, the instances that take longest are not those that also have the largest proofs. Rather, `glucose` takes longest for proofs somewhere between its minimum and maximum proof size. On the other hand, `kissat` has few instances that, counterintuitively, require very small proofs but take a long time to solve. For most instances, `kissat` either outputs a larger proof or takes slightly longer. We suspect that this could be due to `kissat`'s optimizations for specific hard instances, much like the behaviour we observed in `kissat`'s solving time for uniform instances. `kissat` is able to quickly find proofs to most instances, tolerating a larger proof size, while it takes longer for few instances. As dimension increases, both solvers show more and more similar behaviour, with time correlating to proof size.

As mentioned, both solvers behave similarly when it comes to the maximum length of a proof clause. As we discussed in the previous section, the proofs expectably contain very small clauses at low dimensions. At $d = 1$, `kissat`'s proof clauses consist of at most three literals, which could be related to the fact that it solves a 3-SAT instance. Meanwhile, `glucose` also provides very small proof clauses of at most five literals. Until $d = 6$, both solver's proofs require very small clauses and take a bit longer, or they require larger clauses but take a shorter time to solve. At higher dimensions, both solvers start to output longer proof clauses.

Looking at the individual instances, we find that there seems to be a shift from $d = 6$ to $d = 7$. Up to $d = 6$, instances of all densities follow the mentioned pattern of "either large proof or long time to solve", with the distribution of instances by density being different for each solver. However, the instances seem to "trip" from $d = 6$ to $d = 7$ and start correlating their proof size with solver time. This is especially true for the instances around $r_3 \approx 4.267$: Those generally take a longer time to solve, which seems to be connected to proof size. As expected, with increasing dimension, the plots look more and more like those for uniform instances, with instances at $d = 1\,000$ being practically indistinguishable from uniform instances. All of these observations coincide with our finding that instances up to $d = 6$ do not exhibit the well-known "easy-hard-easy" pattern, suggesting that low-dimensional geometric instances behave very unlike uniform instances also with regard to proofs of unsatisfiability.

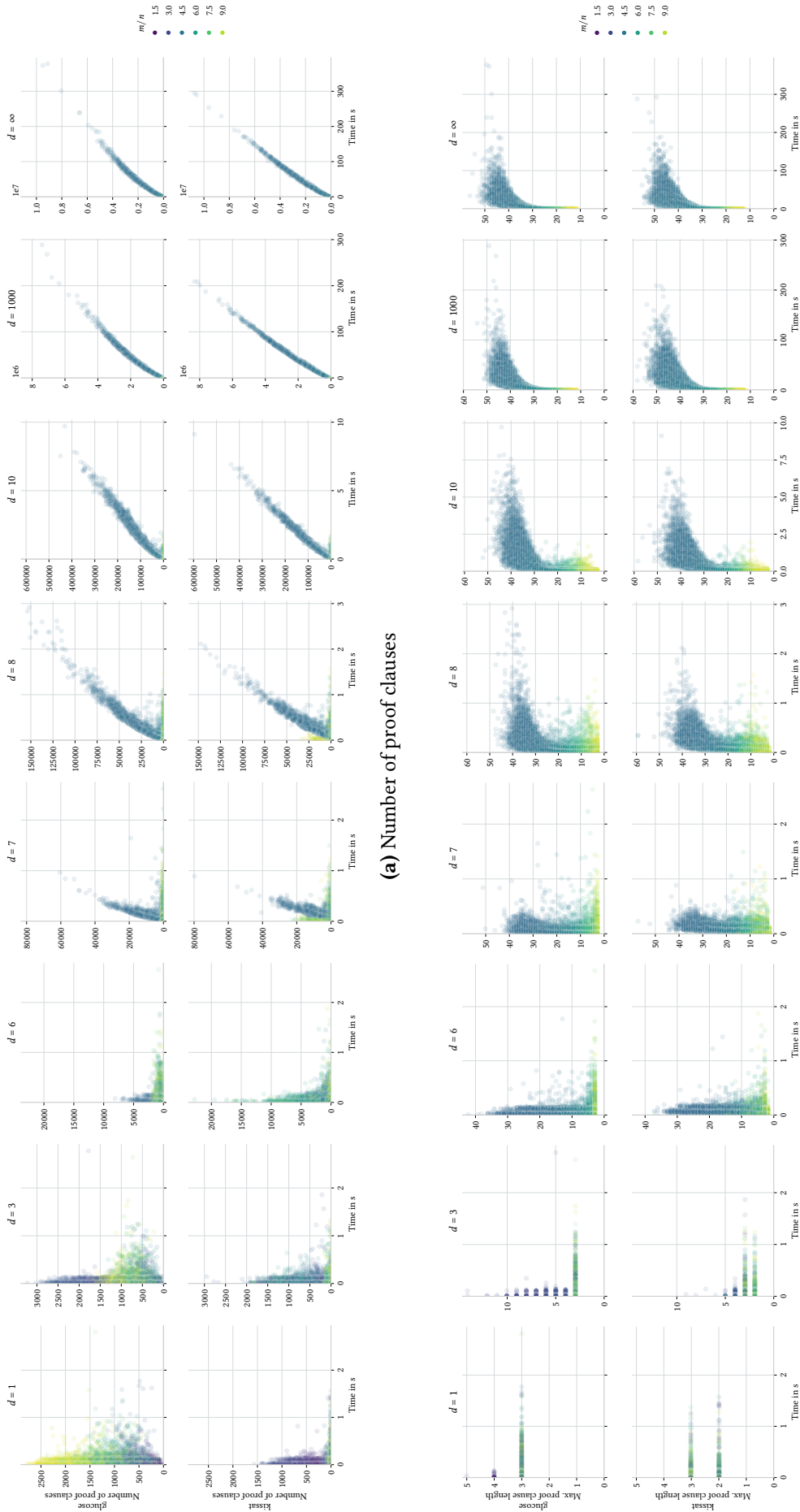

**Figure 4.9:** The relation of time to a proof size metric, with (a) showing the number of clauses and (b) showing the maximum length of a clause. Each row displays values for either glucose or kissat, while each column shows instances at select dimensions $d \in \{1, 3, 6, 7, 8, 10, 1000\}$ up to uniform instances ("$\infty$"). Each instance is represented as a dot, placed according to the time it took to solve and the respective metric for proof size. Colour of a dot indicates the instance's density. Blue-coloured dots are located near the critical density of uniform 3-SAT at $r_3 \approx 4.267$.

## 4.5 Generalisation to Other Instance Sizes

So far, we only looked at 3-SAT instances with a fixed number of variables $n = 300$. To examine the extent to which our results can be generalized to other instance sizes, we generated low-dimensional geometric instances with values of $k \in \{2, 3, 4\}$ and with $n, m \in [20, 1\,000]$. We determined the critical density of those instances, the result is shown in Figure 4.10. As can be seen, our results indeed seem to hold for other instance sizes. Regardless of the number of variables and clause length, the critical density approaches that of the uniform SAT model, increasing slightly with every dimension.

It should be noted that the critical density undergoes a change as the number of variables increases. This is especially true for very small instances with less than 100 variables. Afterwards, the change looks merely sublinear. This is to be expected, since the same effect can be observed for uniform instances. It is especially pronounced for 2-SAT: Although the critical 2-SAT density is proven to be at exactly $r_2 = 1$ [CR92], the critical density of our uniform instances appears to be higher, and it decreases as the number of variables rises due to asymptotic

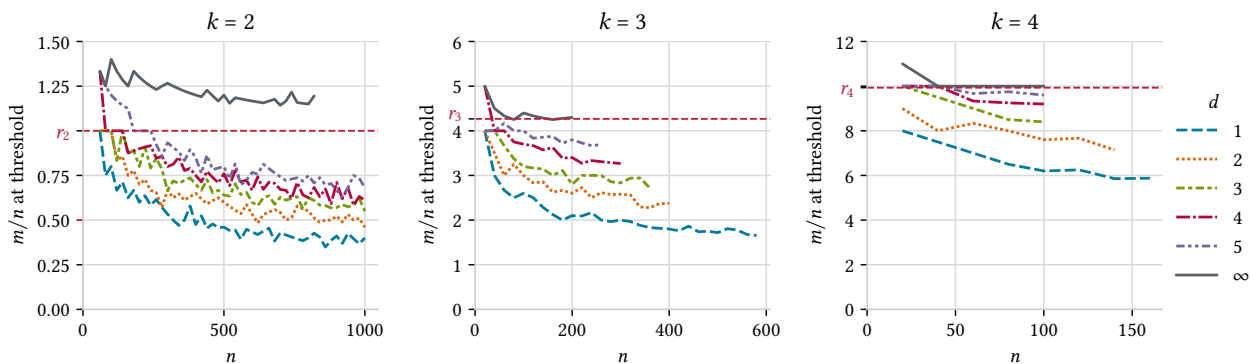

**Figure 4.10**: For clause lengths $k \in \{2, 3, 4\}$, each plot shows the critical density $m/n$ at which $1/2$ of unsatisfiable instances is reached at different numbers of variables $n$. We show results for dimensions $d \in [1, 5]$ as well as uniform instances for comparison ("$\infty$"). Horizontal lines show the critical densities of uniform SAT: $r_1 = 1$ [CR92], and $r_3 \approx 4.267, r_4 \approx 9.931$ as conjectured by [MMZ06].

This was obtained by generating instances with $n, m \in [20, 1\,000]$ in steps of 20, with 50 instances per $m, n$, then finding the density at which the ratio of satisfiable instances is closest to $1/2$ for each $n$.

effects requiring much larger instances. In a separate experiment, we observed that instances at $n \approx 10^7$ are required for the observed critical density to be within 1 % of the expected $r_2$ (these instances are still very tractable since 2-SAT is solvable in polynomial time).

Apart from this, note that at fixed dimension, the critical density of geometric instances increases with clause length. This is in accordance with the critical density of uniform $k$-SAT, which also increases at higher clause lengths.

# 5 Conclusions and Outlook

We have provided an empirical analysis of the geometric SAT model introduced by Giráldez-Cru and Levy [GL17]. We find that dimensionality influences important properties of a geometric instance. As dimension increases, instances from the geometric model seem to converge to those of the uniform SAT model, with high-dimensional instances being practically indistinguishable from uniform instances. Compared to the uniform model, low-dimensional geometric instances appear much more tractable, and they accordingly require smaller proofs by SAT solvers. Additionally, due to the geometry-induced locality, geometric instances are more likely to be unsatisfiable. These observations support the theoretical result by Bläsius *et al.* [Blä+23] that geometric instances are a. a. s. easy to solve.

In accordance with the higher likelihood of being unsatisfiable, the critical density is lower at low-dimensional instances. The satisfiability threshold also appears to be much coarser at low dimensions.

We find that there is a shift in the behaviour of geometric instances from around six dimensions to seven dimensions. Geometric instances at up to six dimensions do not exhibit properties that are typical of uniform instances: Solver time is not influenced much by density, they do not show the "easy-hard-easy" pattern, and proof size does not correlate to solver time. On the other hand, instances behave more and more like uniform instances starting at seven dimensions: Their solver time peaks at the satisfiability threshold and solver time correlates to proof size; though these effects are still less pronounced than in very high-dimensional or uniform instances. Further work is required to gain a detailed understanding of the cause of this shift from six to seven dimensions. Statistical analysis and visualization of the clause-variable graph of an instance could be beneficial in this regard.

Further research with regard to SAT solvers is necessary to explain some of the behaviour we observed, like the plateau in `kissat`'s solver time and the fact that `kissat` is especially fast at the uniform threshold, but slower than `minisat` at other densities and low dimensions. Identifying the responsible design choices in SAT solvers could eventually foster further improvements.

Overall, our findings suggest that geometry provides a promising model for SAT. Further research is required to back this by providing evidence through analysis of industrial instances. The geometric SAT model is capable of reproducing the network structure found in many real-world instances using locality, and it can produce the full range of easy to hard instances by varying the dimension. Paired with a power-law in variable weights, the geometric SAT model could be able to represent a vast range of industrial SAT instances.